\documentclass[psfig]{mn2e}
\usepackage{psfig}
\usepackage{times}

\newcommand{\bd}{\begin{displaymath}}
\newcommand{\ed}{\end{displaymath}}
\newcommand{\be}{\begin{equation}}
\newcommand{\ee}{\end{equation}}

\title{The jet power extracted from a magnetized accretion disc}
\author[X. Cao]
       {Xinwu Cao\\
Shanghai Astronomical Observatory, National Astronomical Observatories,
Chinese Academy of Sciences, \\Shanghai, 200030, China; cxw@center.shao.ac.cn}
           
\date{Accepted . Received ; in original form}

\pagerange{\pageref{firstpage}--\pageref{lastpage}}
\pubyear{}

\begin{document}

\maketitle
\label{firstpage}

\begin{abstract}
We consider the power of a relativistic jet accelerated by the 
magnetic field of an accretion disc. It is found that the power
extracted from the disc is mainly determined by the field strength and
configuration of the field far from the disc. Comparing 
with the power extracted from a rotating black hole, we find that the jet
power extracted from a disc can dominate over that from
the rotating black hole. But in some cases, the jet power extracted from
a rapidly rotating hole can be more important than that from the disc
even if the poloidal field threading the hole is not significantly
larger than that threading the inner edge of the disc. The results
imply that the radio-loudness of quasars may be governed by its
accretion rate which might be regulated by the central black hole
mass. It is proposed that the different disc
field generation mechanisms might be tested against observations of
radio-loud quasars if their black hole masses are available. 
\end{abstract}

\begin{keywords}
accretion, accretion discs--galaxies: jets--galaxies: nuclei
\end{keywords}

\section{Introduction}

The current most favoured models of powering active galactic nuclei (AGNs)
involve gas accretion onto a massive black hole, though the details
are still unclear. Relativistic jets have been observed in many radio-loud
AGNs and are believed to be formed very close to the black holes.
The spin energy of a black hole might be extracted to fuel the jet
by magnetic fields supported by a surrounding accretion disc
(Blandford \& Znajek 1977; Blandford-Znajek process). This process has
widely been believed to be the major mechanism that powers radio jets
in AGNs (Begelman, Blandford \& Rees 1984; Rees et al. 1982;
Wilson \& Colbert 1995; Moderski \& Sikora 1996).
However, Ghosh \& Abramowicz (1997) doubted the importance of the
Blandford-Znajek process. For a black hole of a given mass and
angular momentum, the strength of the Blandford-Znajek process depends
crucially on the strength of the poloidal field threading the horizon
of the hole. They argued that the strength of the field threading
a black hole has been overestimated. The magnetic field threading a
hole should be maintained by the currents situated in the inner region
of a surrounding accretion disc. Livio, Ogilvie \& Pringle (1999;
hereafter L99) re-investigated the problem and pointed out that even
the calculations of Ghosh \& Abramowicz (1997) have overestimated the
power of the Blandford-Znajek process. Ghosh \& Abramowicz (1997)
have overestimated the strength of the large-scale field threading the
inner region of an accretion disc, and then the power of the
Blandford-Znajek process. They argued that the jet power extracted
from an accretion disc dominates over the power extracted by the
Blandford-Znajek process (L99).

L99 estimated the maximal electromagnetic output from an accretion disc.
In this case, the toroidal field component is of the same order of the
poloidal field component at the disc surface due to the instability of
a predominantly toroidal field (Biskamp 1993; Livio \& Pringle 1992). Apart
from the strength of the field threading a disc, the acceleration of the 
jet is also governed by the magnetic field configuration and the
structure of the disc (Shu 1991; Ogilivie \& Livio 1998, 2000;
Cao \& Spruit 2001; for relativistic jets near rotating black holes, see
Koide, Shibata \& Kudoh 1999; Koide et al. 2000).
It is unclear how much power can actually be
extracted from a magnetized accretion disc without considering its field
configuration. In this work, we extend L99's work to explore some
factors that affect the power of a relativistic jet extracted
from an accretion disc.

\section{The power of a jet extracted from an accretion disc}

For a relativistic jet accelerated by the magnetic field of the disc, the
Alfv{\'e}n velocity is (Michel 1969; Camenzind 1986)

\be
v_{\rm A}={\frac {{B_{\rm p}^{\rm A}}}
{(4\pi\rho_{\rm A}\gamma_{\rm j})^{1/2}}},
\ee
where $B_{\rm p}^{\rm A}$ and $\rho_{\rm A}$ are the poloidal field strength
and density of the jet at Alfv{\'e}n point, and $\gamma_{\rm j}$ is the
Lorentz factor of the bulk motion of the jet. Equation (1) is a good
approximation, since the Alfv{\'e}n point is usually not very close
to the rotating black hole (Camenzind 1986). The value of the
Alfv{\'e}n velocity is of the order
$v_{\rm A}\sim R_{\rm A}\Omega(R_{\rm d})$, where $R_{\rm d}$ is the
radius of the field footpoint at the disc surface, and $R_{\rm A}$ is
the Alfv{\'e}n radius of the jet along the field line.

Mass and magnetic flux conservation along a magnetic field line
requires

\be
{\frac { {\dot m}_{\rm jet} }{B_{\rm pd}} }\simeq
{\frac {\rho_{\rm A}v_{\rm A} }{B_{\rm p}^{\rm A}} },
\ee
where ${\dot m}_{\rm jet}$ is the mass loss rate in the jet from 
unit surface area of the disc, and $B_{\rm pd}$ is the poloidal field
strength at the disc surface.

Since the Alfv{\'e}n radius $R_{\rm A}$ depends on the global field
configuration, not only the strength of the field at the disc surface,
we assume the magnetic field far from the disc surface to be roughly
self-similar:

\be
B_{\rm p}^{\rm A} \sim B_{\rm pd}\left( {\frac {R_{\rm A}}{R_{\rm d}}}
\right)^{-\alpha},
\ee
where the self-similar index $\alpha>1$ (Blandford \& Payne 1982;
Lubow, Papaloizou \& Pringle 1994). A self-similar index $\alpha=4$
is adopted in the calculations of Lubow et al. (1994). 

The Lorentz factor of the jet is

\be
\gamma_{\rm j}\simeq 
\left[1-\left({\frac {v_{\rm A}} {c}}\right)^2\right]^{-{1\over 2}}.
\ee

Combining Eqs (1)-(4), the mass loss rate in the jet from unit surface 
area of the disc is

\be
{\dot m}_{\rm jet}=
{\frac {B_{\rm pd}^2}{4\pi c}}
\left(
{\frac {R_{\rm d}\Omega(R_{\rm d})} {c}} \right)^{\alpha}
{\frac {\gamma_{\rm j}^\alpha}
{(\gamma_{\rm j}^2-1)^{\frac {1-\alpha}{2}}}}.
\ee

The power of the jet accelerated by the magnetized disc is

\bd
L_{\rm d}=2\pi\int (\gamma_{\rm j}-1){\dot m}_{\rm jet}c^{2}
R_{\rm d}dR_{\rm d}
\ed
\be
=2\pi c\int
{\frac {B_{\rm pd}^2}{4\pi}}
\left(
{\frac {R_{\rm d}\Omega(R_{\rm d})} {c}} \right)^{\alpha}
f(\alpha,~\gamma_{\rm j})R_{\rm d}dR_{\rm d}, 
\ee
where the function $f(\alpha,~\gamma_{\rm j})$ is given by

\be
f(\alpha,~\gamma_{\rm j})=
{\frac {(\gamma_{\rm j}-1)\gamma_{\rm j}^\alpha}
{(\gamma_{\rm j}^2-1)^{\frac {1-\alpha}{2}}}}.
\ee
For a given $\gamma_{\rm j}$, Eq. (6) reduces to

\be
L_{\rm d}
=2\pi c f(\alpha,~\gamma_{\rm j})
\int
{\frac {B_{\rm pd}^2}{4\pi}}
\left(
{\frac {R_{\rm d}\Omega(R_{\rm d})} {c}} \right)
R_{\rm d}dR_{\rm d},  
\ee
in the case of $\alpha=1$. The power of the jet $L_{\rm d}$
decreases with the self-similar index $\alpha$, since the term
$R_{\rm d}\Omega(R_{\rm d})/c$ in Eq. (6) is always less than unity.
Thus, Eq. (8) gives an upper limit on the power of a
relativistic jet for given field strength at the disc surface and
Lorentz factor $\gamma_{\rm j}$ of the jet ($\gamma_{\rm j}\gg 1$;
see Fig. 1). Note that Eq. (8) is almost same
as Eq. (6) in L99 for the maximal electromagnetic output from the disc,
except an additional term of $f(\alpha,~\gamma_{\rm j})$ appears here.
Here we have obtained a similar conclusion as that in L99, but in a rather
different way.

The toroidal field strength at the disc surface is

\be
B_{\phi\rm d}={\frac {4\pi}{B_{\rm pd}}}
{\frac {\dot{m}_{\rm jet}c^{2} }
{R_{\rm d}\Omega(R_{\rm d})} }
(\gamma_{\rm j}-1).
\ee

Substituting Eq. (5) into Eq. (9), we have

\be
{\frac {B_{\phi\rm d}}{B_{\rm pd}}}=
\left( {\frac {R_{\rm d}\Omega(R_{\rm d})} {c}}\right)^{\alpha-1}
f(\alpha,~\gamma_{\rm j}).
\ee
It is obvious that the ratio $B_{\phi\rm d}/B_{\rm pd}$ at the given
radius $R_{\rm d}$ is determined by the self-similar index
$\alpha$ and the Lorentz factor $\gamma_{\rm j}$, i.e., the ratio
only depends on the jet properties, not on the disc properties.

\subsection{The magnetic field strength at the disc surface}

In order to complete the estimate on the power of the jet
extracted from an accretion disc, we need to know the field strength
at the disc surface. The origin of the ordered field that is assumed
to thread the disc is still unclear. The strength of the field at the
disc surface is usually assumed to scale with the pressure of the disc, as
done in Ghosh \& Abramowicz (1997). However, L99 have estimated the
strength of the large-scale field threading the disc 
based on the results of some numerical simulations on dynamo mechanisms in
accretion discs (Tout \& Pringle 1996; Amitage 1998; Romanova et al. 1998).
In this work, we mainly follow L99's estimate on the field strength at
the disc surface. The physical implications of these two different
estimates on the field strength at the disc surface will be discussed
in Sects. 4 and 5.

As L99, the strength of the magnetic field produced by dynamo
processes in the disc is given by

\be
{\frac {B_{\rm dynamo}^2}{4\pi}} \sim
{\frac {W}{2 H}},
\ee
where $W$ is the integrated shear stress of the disc, and $H$ is the
scale-height of the disc. For a relativistic accretion disc, the
integrated shear stress is given by Eq. (5.6.14a) in Novikov \&
Thorne (1973, hereafter NT73). Equation (9) can be re-written as

\be
B_{\rm dynamo}=3.56\times 10^8 r_{\rm d}^{-3/4}m^{-1/2}
A^{-1}BE^{1/2} {\rm gauss},
\ee
where the dimensionless quantities are defined by

\be
r_{\rm d}={\frac {R_{\rm d}}{R_{\rm G}}},~~~
R_{\rm G}={\frac {GM}{c^2} },~~~
m={\frac {M}{M_{\odot}} },
\ee
and $A$, $B$ and $E$ are general relativistic correction factors
defined in NT73.

In standard accretion disc models, the angular velocity of the matter
in the disc is usually very close to Keplerian velocity. For a
relativistic accretion disc surrounding a rotating black hole,
the Keplerian angular velocity is given by

\be
\Omega(r_{\rm d})
={\frac {M^{1/2}}{r_{\rm d}^{3/2}+a} },
\ee
where $a$ is dimensionless specific angular momentum of a rotating
black hole. 

The large-scale field can be produced from the small-scale field created
by dynamo processes as $B(\lambda)\propto \lambda^{-1}$ for the idealized
case, where $\lambda$ is the length scale of the field (Tout \& Pringle
1996; Romanova et al. 1998). The large-scale field threading the disc
is related with the field produced by dynamo processes approximately
by (L99)

\be
B_{\rm pd}\sim \left( {\frac {H}{R_{\rm d}}} \right)_{\rm max}
B_{\rm dynamo}. 
\ee
The dimensionless scale-height of a disc $h=H/R_{\rm d}$ is in
principle a function of $R_{\rm d}$, and it reaches a maximal value
in the inner region of the disc (Laor \& Netzer 1989). We
adopt the maximal value of $H/R_{\rm d}$ here.

The scale-height of the disc $H/R_{\rm d}$ is given by (Laor \& Netzer
1989)

\be
{\frac {H}{R_{\rm d}} } =26.345\dot{m}_{\rm acc}r_{\rm d}^{-1}c_{2},
\ee
where the coefficient $c_{2}$ is defined in NT73. The dimensionless
accretion rate is defined by (Laor \& Netzer 1989)

\be
\dot{m}_{\rm acc}={\frac {\dot{M}_{\rm acc}}{1.626\times 10^{15}
m~~{\rm g~s^{-1}} } }.
\ee

We use Eqs. (12) and (15), the power of the jet
accelerated from a magnetized disc is available by integrating Eq. (6),
if some parameters: $m$, $\dot{m}_{\rm acc}$, $a$, $\alpha$ and
$\gamma_{\rm j}$ are specified.

\section{The relative importance of the Blandford-Znajek process}

As discussed in L99, the power extracted from a rotating black hole
by the Blandford-Znajek process is determined by the hole mass $M$, the
spin of the hole $a$, and the strength of the poloidal field threading
the horizon of a rotating hole $B_{\rm ph}$:

\be
L_{\rm BZ}({\rm max})\sim
{\frac {B_{\rm ph}^{2}}{4\pi}}
\pi R_{\rm h}^{2}a^{2}c,
\ee
where $R_{\rm h}$ is the radius of the black hole horizon. The strength
of the poloidal field threading the hole is a crucial factor determining
the power extracted from the hole. As argued in Ghosh \& Abramowicz (1997)
and L99, the field threading the black hole is generated by the currents
situated in the disc rather than in the hole. Thus, it is realized
that the field threading the hole should not be significantly larger
than the field threading the inner edge of the disc  Here we use a
parameter $\zeta$ to relate $B_{\rm ph}$ with $B_{\rm pd}(r_{in})$:

\be
B_{\rm ph}=\zeta B_{\rm pd}(r_{\rm in}),
\ee
where $\zeta\ge 1$, $r_{\rm in}$ is the radius of the inner edge of the
disc. For a standard accretion disc, $r_{\rm in}=r_{\rm ms}$, where
$r_{\rm ms}$ is the radius of the minimal stable circular orbit.

The Numerical simulations on the Blandford-Znajek process show that it can
operate near its maximum power output (Komissarov 2001).  
Using Eqs. (6), (18) and (19), we can compare the jet power extracted from
a magnetized disc with that by the Blandford-Znajek process. It is
worth noting that the ratio of the jet power extracted from the disc to that
from a rotating black hole is only determined by the parameters
$\alpha$, $\gamma_{\rm j}$ and $\zeta$. The relative importance of these
two processes is independent of the mechanism for producing
the disc field.

\section{Results}

We plot the function $f(\alpha,~\gamma_{\rm j})$ varying with
$\gamma_{\rm j}$ for different values of $\alpha$ in Fig. 1. The final
velocity of a magnetically driven jet depends on the strength and global
configuration of the field threading the disc, and the structure of
the disc that provides boundary conditions for the jet
(Shu 1991; Ogilivie \& Livio 1998, 2000; Cao \& Spruit 1994, 2001).
In this work, we focus on relativistic jets which are widely observed
in AGNs. To avoid being involved in the complexity of the jet acceleration
problem, we use the Lorentz factor $\gamma_{\rm j}$ of a 
magnetically accelerated jet as a parameter to describe the problem.
Now the problem is described by the parameters $m$, $a$, $\dot{m}$,
$\alpha$ and $\gamma_{\rm j}$.

The mass loss rate in the jet from unit surface area of the
disc is plotted in Figs. 2 and 3 for the non-rotating and rapidly
rotating black holes respectively.  We plot the total mass loss rate
in the jet and the power of the jet as functions of $\gamma_{\rm j}$
for different values of $\alpha$ in Figs. 4 and 5 respectively.
We calculate the ratio of the toroidal field strength to the poloidal
one at the disc surface using Eq. (10), and the results are present
in Fig. 6.

Using Eq. (18), we can estimate the maximal power output from a rotating
black hole by the Blandford-Znajek process.
The power outputs as functions of $a$ are present in Fig. 7
for different values of $\alpha$ and $\zeta$. We plot the power outputs
varying with accretion rate $\dot{m}_{\rm acc}$ based on L99's estimate
of the disc field strength in Fig. 8. Finally, using Eqs. (6), (18) and
(19), we plot the relative importance of these two power extract
processes in the parameter space ($\alpha$-$a$) in Fig. 9.

\subsection{The ratio of jet power to accretion luminosity}

The power of the jet extracted from a disc or a rotating black hole 
depends mainly on the field strength at the disc surface, since 
the field threading the hole is generated by the currents situated
in the disc. The ratio of jet power to accretion luminosity is therefore
related with the mechanism for generating the disc field.

In the case of the large-scale field being produced from the small-scale
field created by dynamo processes, the large-scale field strength is
estimated by Eq. (15). As we know, $H/R_{\rm d}\propto \dot{m}_{\rm acc}$
and $L_{\rm acc} \propto \dot{m}_{\rm acc}m$, then we obtain a linear
relation between $\dot{m}_{\rm acc}$  and $L_{\rm jet}/L_{\rm acc}$
by using Eqs. (6) (see Fig. 8):

\be
L_{\rm jet}/L_{\rm acc}\propto \dot{m}_{\rm acc},
\ee
which is independent of the black hole mass. Note that the strength of the field
threading the hole is related with that threading the disc (see Eqs. (18)
and (19)), the relation (20) is also valid for jet power extracted from
the rotating hole. 

The similar analysis can be performed for the case that the magnetic field
strength scales with the pressure of the disc.  
The strength of the field at the disc surface is

\be
B_{\rm pd}(r_{\rm d})
\propto r_{\rm d}^{-3/4}m^{-1/2}
A^{-1}BE^{1/2}, 
\ee
which is in dependent of the accretion rate (NT73).

From Eqs. (6), (18) and (21), we obtain a relation (Ghosh \& Abramowicz
1997)

\be
L_{\rm jet}/L_{\rm acc}\propto \dot{m}_{\rm acc}^{-1}, 
\ee
which holds for a jet accelerated either from a disc or a rotating hole.
It does not depend on the black hole mass either.

\begin{figure}
\centerline{\psfig{figure=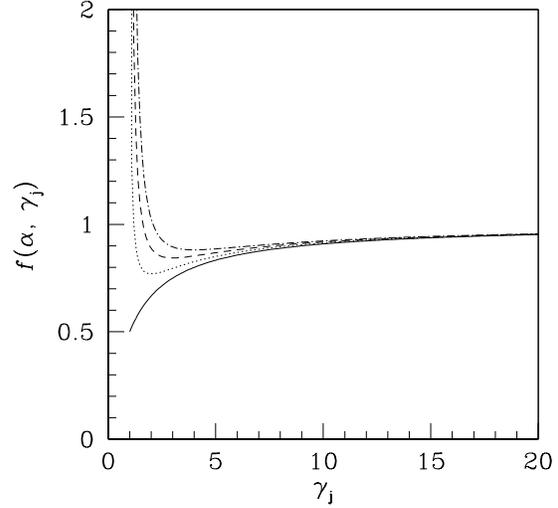,width=7.5cm,height=7.5cm}}
\caption{The function $f(\alpha,~\gamma_{\rm j})$ varies with
$\gamma_{\rm j}$ for different values
of $\alpha$: $\alpha=1$(solid line), 2(dotted), 3(dashed)
and 4(dash-dotted).}
\end{figure}

\begin{figure}
\centerline{\psfig{figure=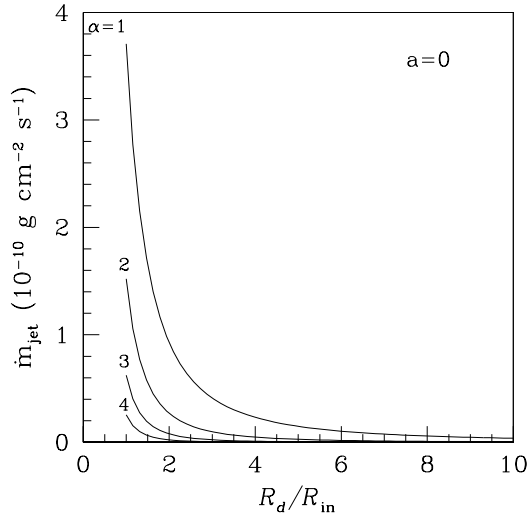,width=7.5cm,height=7.5cm}}
\caption{The mass loss rate in the jet from unit surface area of the
disc surrounding a non-rotating black hole for different values of $\alpha$.
The hole mass $m=10^9$, $\gamma_{\rm j}=10$ and $\dot{m}_{\rm acc}=0.05$
are adopted. }
\end{figure}

\begin{figure}
\centerline{\psfig{figure=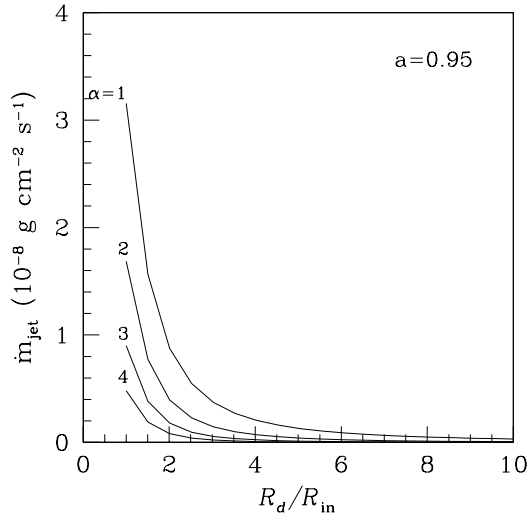,width=7.5cm,height=7.5cm}}
\caption{Same as Fig. 2, but for a rapidly rotating black hole:
$a=0.95$.  }
\end{figure}

\begin{figure}
\centerline{\psfig{figure=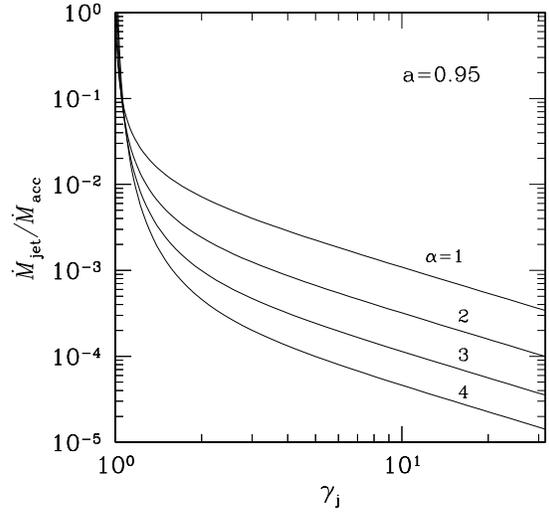,width=7.5cm,height=7.5cm}}
\caption{ The total mass loss rate in the jet as functions of
$\gamma_{\rm j}$ for different values of $\alpha$ for the field.
The black hole spin parameter $a=0.95$ and $\dot{m}_{\rm acc}=0.05$
are adopted.}
\end{figure}

\begin{figure}
\centerline{\psfig{figure=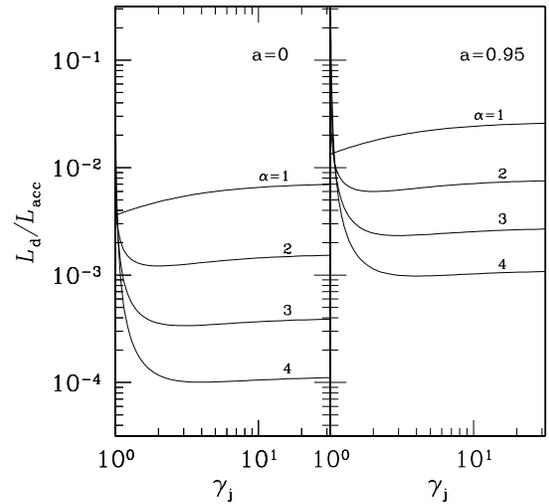,width=7.5cm,height=7.5cm}}
\caption{ The power of the jet accelerated
by a magnetized disc as functions of $\gamma_{\rm j}$ for
different values of $\alpha$. The left panel is for
a non-rotating black hole, while the right panel is for a rapidly
rotating black hole: $a=0.95$. The accretion rate $\dot{m}_{\rm acc}
=0.05$ is adopted.}
\end{figure}

\begin{figure}
\centerline{\psfig{figure=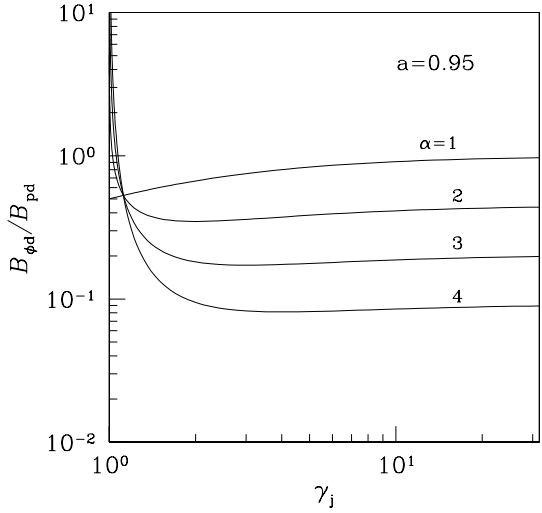,width=7.5cm,height=7.5cm}}
\caption{The ratio of the toroidal field strength to the poloidal
one at the disc surface as functions of $\gamma_{\rm j}$ for different
values of $\alpha$ at $R_{\rm d}=2R_{\rm in}$. The hole spin
parameter $a=0.95$ is adopted.}
\end{figure}

\begin{figure}
\centerline{\psfig{figure=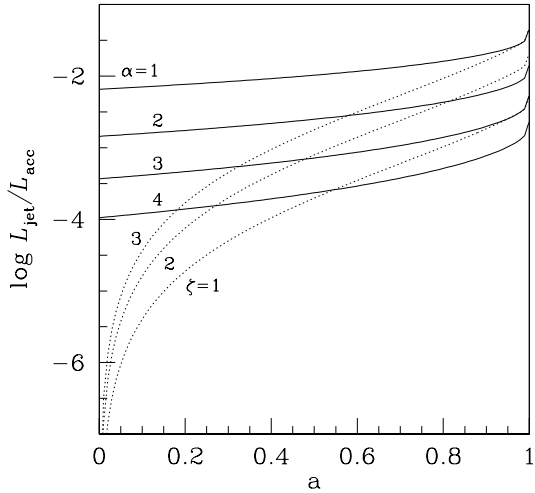,width=7.5cm,height=7.5cm}}
\caption{The power outputs as functions of $a$ 
for different values of $\alpha$ and $\zeta$. The solid lines
represent the power output from an accretion disc, while the
dotted lines represent the power output of the Blandford-Znajek
process. The accretion rate $\dot{m}_{\rm acc}=0.05$ is adopted. }
\end{figure}

\begin{figure}
\centerline{\psfig{figure=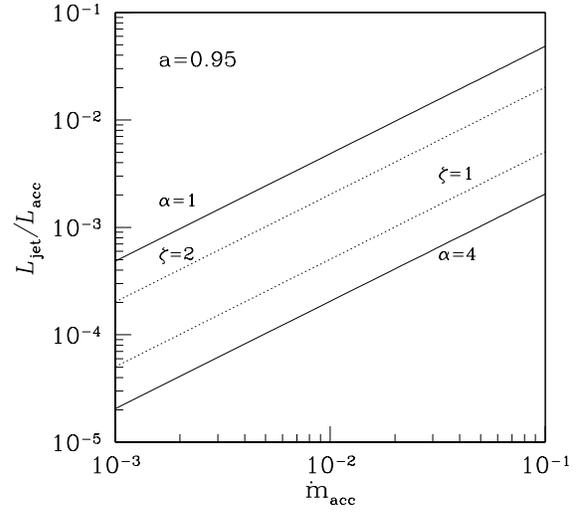,width=7.5cm,height=7.5cm}}
\caption{The power outputs as functions of $\dot{m}_{\rm acc}$ 
for different values of $\alpha$ and $\zeta$. The solid lines
represent the power output from an accretion disc, while the
dotted lines represent the power output of the Blandford-Znajek
process. The black hole spin parameter $a=0.95$ is adopted. }
\end{figure}

\begin{figure}
\centerline{\psfig{figure=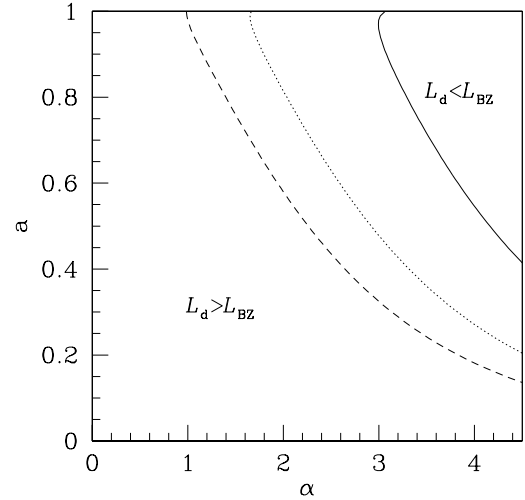,width=7.5cm,height=7.5cm}}
\caption{The relative importance of two power extract processes in
parameter space ($\alpha$-$a$) for different values of $\zeta$:
$\zeta=1$(solid line), 2(dotted) and 3(dashed). $\gamma_{\rm j}=10$ is
adopted. }
\end{figure}

\section{Discussion of the results}

Figure 1 shows that the function $f(\alpha,~\gamma_{\rm j})$ varies
slowly with $\gamma_{\rm j}$ except for $\gamma_{\rm}\rightarrow 1$.
The function $f(\alpha,~\gamma_{\rm j})$ is almost a constant while
$\gamma_{\rm j}>5$. From Eq. (6), it is then found that the power 
of a relativistic jet accelerated by a magnetized accretion disc
is insensitive to its Lorentz factor $\gamma_{\rm j}$. This is
confirmed by the results plotted in Fig. 5. The jet power varies
with $\gamma_{\rm j}$ in a range less than 10 \% for $\gamma_{\rm j}>5$.
It implies that relativistic jets with different Lorentz factors
$\gamma_{\rm j}$ accelerated by magnetized discs have almost same
jet power, if all other physical parameters are fixed.
We can therefore adopt a typical value of $\gamma_{\rm j}$ in our
calculations, which will not change the main results on jet power, 
though the terminal velocity of the magnetically driven jet could be
given in principle only after the structure of the disc and its
field strength and configuration are supplied.  
This has simplified our present investigation, since we are interested
in the power of relativistic jets extracted from discs.

The power of a jet decreases with the self-similar index $\alpha$
(see Fig. 5). A large $\alpha$ implies that the poloidal field
strength decays rapidly with radius along a field line.
From Fig. 4, we see that less mass is loaded into the jet in the
case of  a larger $\alpha$, and the power extracted from the disc is
then reduced. The power of the jet is sensitive to the field configuration
far from the disc surface.

We find that the mass loss rate in the jet from unit surface area of
the disc decreases with radius (see Figs. 2 and 3). Almost all mass
carried by the jet is from the region of the disc:
$R_{\rm d}<5~R_{\rm in}$, i.e., jets
are mainly accelerated from the inner region of the disc where most
gravitational energy of accretion matter is released. The total mass
loss rate in a relativistic jet can be neglected compared with the
accretion rate of the disc (usually
$\dot{M}_{\rm jet}<10^{-2}\dot{M}_{\rm acc}$, see Fig. 4). A jet with
high mass loss rate results in a low terminal velocity of the jet and
wound-up field lines (Figs. 4 and 6), which is consistent with the
results of Cao \& Spruit (1994). As pointed out in
Sect. 2, our calculation of the jet power extracted from a disc can
reproduce L99's result while $\alpha=1$ is adopted (see Eq. (8)). In
L99's calculation, $B_{\phi\rm d}=B_{\rm pd}$ at the disc surface is
assumed, so it is not surprised to find that $B_{\phi\rm d}\simeq
B_{\rm pd}$ in the case of $\alpha=1$ in our calculations (see Fig. 6).
For a large $\alpha$, the ratio $B_{\phi\rm d}/B_{\rm pd}$ is low
(compare the results in Fig. 6 for different values of $\alpha$). 

In our present calculations, the field configuration far from the
disc is described by the self-similar index $\alpha$. In the case of $\alpha=1$,
it is found in Fig. 7 that the power extracted from the disc always
dominates over the maximal power of the Blandford-Znajek process, if
$\zeta<3$, i.e., the strength of the poloidal field threading the hole
is less than three times of that threading the inner edge of the disc.
If $\zeta=1$ is assumed, i.e., $B_{\rm ph}=B_{\rm pd}(R_{\rm in})$,
$L_{\rm d}$ is an order of magnitude larger than $L_{\rm BZ}$ even
for an extreme Kerr black hole. Thus the conclusion given by L99 that
the maximal power extracted from a disc dominates over that from a
rapidly rotating black hole is confirmed, while our calculations are
not limited to the extreme case $\alpha=1$. We present the results in
Fig. 7 to compare the power outputs by these two processes for
different values of $\alpha$ and $\zeta$. It is found that the
Blandford-Znajek process would be important for a large $\alpha$
or/and $\zeta$. From Eqs. (6) and (18), we find that this conclusion
is independent of the mechanism for producing disc fields. 
The conclusion is also independent of the hole mass $m$
and accretion rate $\dot{m}_{\rm acc}$, since the hole field strength
is only governed by the disc field strength.

If the large-scale field can be produced from the small-scale field
created by dynamo processes as done by L99,
we find that linear relations are present:
$L_{\rm jet}/L_{\rm acc}\propto \dot{m}_{\rm acc}$, for a jet accelerated
either from a disc or a rotating hole. (see Fig. 8 and Eq. (20)).
This relation is quite different from the situation for the field that
is assumed to scale with the pressure in the disc, where  
$L_{\rm jet}/L_{\rm acc}\propto \dot{m}_{\rm acc}^{-1}$ is present.

The ratio $L_{\rm jet}/L_{\rm acc}$ can be related with an observational
quantity of quasars: radio-loudness $R$, which is defined as
$R\equiv f_{\nu}{\rm (5GHz)}/f_{\nu}(\rm{4400\AA})$. 
The fact that the ratio $L_{\rm jet}/L_{\rm acc}$ is
independent of the black hole mass may imply that the radio-loudness $R$
is not related directly with the black hole mass. The
radio-loudness $R$ may probably governed by the accretion rate.
The accretion rate is then regulated by the central black hole mass
in some way (Yi 1996; Salucci et al. 1999; Haiman \& Menou 2000;
Kauffmann \& Haehnelt 2000). The relations found between
radio-loudness and the central black hole masses in quasars
(McLure et al. 1999; Laor 2000; Gu, Cao \& Jiang 2001; Ho 2002)
may reflect intrinsic relations between radio-loudness and
accretion rate which is regulated by the black hole mass (Boroson 2002).
The different relations between $L_{\rm jet}/L_{\rm acc}$ and
$\dot{m}_{\rm acc}$ are expected for different mechanisms of
producing the disc field. The different disc field generation mechanisms
can therefore be tested against observations of radio-loud quasars if
their black hole masses are available. However, whether the jet power
is extracted from a disc or a rotating black hole cannot be tested in
this way, since the calculations predict the same relation between
$L_{\rm jet}/L_{\rm acc}$  and $\dot{m}_{\rm acc}$ for these two jet
acceleration mechanisms.

In Fig. 9, we compare the relative importance of these two power
extract processes in the parameter space ($\alpha-a$).
Only for those cases the strength of the
poloidal field of the disc decays slowly with radius along the field
line (small $\alpha$), the power extracted from the disc dominates over
that extracted by the Blandford-Znajek process even for a rapidly
rotating hole. In the right-upper
corner of the figure, i.e., the case with a large $\alpha$ and $a$,
the power extracted from the disc is greater than the power of the
Blandford-Znajek process. Otherwise, the power of the Blandford-Znajek
process can be neglected compared with that from the disc.

The acceleration of the jet is governed by the structure of the disc and
the strength and configuration of the field threading the disc. The
physical properties of the jet, such as the power and terminal velocity
of the jet, can be determined if all these physical factors of the disc
and its field are known. Instead of solving a set of MHD equations
describing the disc-jet system, a self-similar index $\alpha$ is employed to describe
the field far above the disc surface in present work. The physics of how
the jet acceleration is governed by the detailed structure of the disc
near a rapidly rotating black hole has not been included in this work.
It would be the main limitation of the present work.

\section*{Acknowledgments}
I thank the anonymous referee for his/her helpful suggestions that
improved the presentation of the paper.
The support from the NSFC
(No. 10173016) and NKBRSF (No. G1999075403) is gratefully acknowledged.

{}

\end{document}